\documentstyle[11pt,aaspp4]{article}

\def\urp{\hbox{$^{2}\Pi_{1/2}$}}
\def\ulp{\hbox{$^{2}\Pi_{3/2}$}}
\def\E#1{\hbox{$10^{#1}$}}
\def\about  {\hbox{$\sim$}}
\def\x      {\hbox{$\times$}}
\def\kms    {\hbox{km\,s$^{-1}$}}
\def\NOH    {\hbox{$N_{\rm OH}$}}

\def\cc     {\hbox{cm$^{-3}$}}
\def\cs     {\hbox{cm$^{-2}$}}
\def\la     {\hbox{$\lesssim$}}
\def\ga     {\hbox{$\gtrsim$}}

\begin{document}

\rightline{To appear in ApJ 511, January 20, 1999}

\title{OH 1720 MHz Masers in Supernova Remnants --- C-Shock Indicators}

\author{Phil Lockett\altaffilmark{1}, Eric Gauthier\altaffilmark{1,2},
        and Moshe Elitzur\altaffilmark{3}}

\altaffiltext{1}{Centre College, 600 West Walnut Street, Danville, KY 40422;
lockett@centre.edu}

\altaffiltext{2}{Department of Physics \& Astronomy, The Johns Hopkins
University, 3400 North Charles Street, Baltimore, MD 21218;
gauthier@pha.jhu.edu}

\altaffiltext{3}{Department of Physics and Astronomy, University of Kentucky,
Lexington, KY 40506-0055; moshe@pa.uky.edu}

\begin{abstract}
Recent observations show that the OH 1720 MHz maser is a powerful probe of the
shocked region where a supernova remnant strikes a molecular cloud. We perform
a thorough study of the pumping of this maser and find tight constraints on the
physical conditions needed for its production. The presence of the maser
implies moderate temperatures (50 -- 125 K) and densities (\about\ \E5 \cc),
and OH column densities of order \E{16} cm$^{-2}$. We show that these
conditions can exist only if the shocks are of C-type. J-shocks fail by such a
wide margin that the presence of this maser could become the most powerful
indicator of C-shocks. These conditions also mean that the 1720 MHz maser will
be inherently weak compared to the other ground state OH masers. All the model
predictions are in good agreement with the observations.
\end{abstract}

\section{Introduction}

The ground state maser lines of the OH molecule have long been important
sources of information about the regions in which they are formed. The main
lines at 1665 and 1667 MHz and the satellite line at 1612 MHz have been widely
observed in star forming regions and the envelopes of late-type stars. The
satellite line at 1720 MHz has been the least commonly observed and has
received the least attention until recently. However, the 1720 MHz maser is now
known to be an important indicator of interactions between supernova remnants
and molecular clouds. This connection was first suggested by Goss \& Robinson
(1968), who noted a new class of 1720 MHz masers in the supernova remnants
(SNR) W28 and W44. They detected strong 1720 MHz emission, while the other
ground state lines were in absorption. Little additional study of these masers
was performed until recently, when the connection between 1720 MHz maser
emission and SNR's was clearly established by Frail, Goss \& Slysh (1994). In a
followup survey, 1720 MHz maser emission was detected in 17 out of 160 SNR's
(Green et al. 1997). Another source of 1720 MHz maser emission has also been
discovered near the galactic center (Yusef-Zadeh et al. 1996). All these masers
appear to be the result of shock waves propagating perpendicular to the line of
sight, hitting adjacent molecular clouds. In their comprehensive study, Frail
et al. (1996) note that the maser velocities are usually close to those of the
surrounding gas. This finding is corroborated by Claussen et al.\ (1997), who
also find in the particular case of IC443 that the masers are confined to one
clump of the SNR. Earlier molecular line studies by van Dishoeck, Jansen, \&
Phillips (1993) showed that, in that particular clump, the shock is propagating
perpendicular to the line of sight. Therefore, as could be expected, the maser
emission is strongest parallel to the shock front, the direction of largest
velocity coherence along the line of sight.

The molecular line studies of van Dishoeck et al.\ (1993) also find that in
IC443 the pre-shock temperature is \about\ 15 K and density \about\ \E4\ \cc.
The corresponding postshock values are \about\ 80 K and \about\ \E5\ \cc,
respectively. These postshock measurements are supported by the recent H$_2$CO
and CO observations of Frail \& Mitchell (1998), who determined a temperature
$\sim$ 80 K and densities of \E{4} -- \E{5} \cc \ in their study of W28, W44
and 3C391. Estimates of the post-shock OH column density in W28 range from
\E{16} to \E{17} cm$^{-2}$ (Frail et al. 1994).  Finally, Yusef-Zadeh et al.\
(1996), Claussen et al.\ (1997) and Koralesky et al.\ (1998) detected strong
polarization and inferred magnetic fields up to a few milligauss in these maser
regions.

The basic pumping mechanism for the 1720 MHz masers was identified long ago
(Elitzur 1976; E76 hereafter) as collisional excitations at moderate
temperatures ($T$ \la~200 K), densities ($n$ \la~\E5\ \cc) and OH column
densities (\NOH\ \about\ \E{16}--\E{17} cm$^{-2}$). Here we refine these old
calculations with the latest collision rates and use the constraints imposed by
the pumping requirements to construct a detailed model for the masers around
SNR's. In section II we discuss the collisional pumping of the 1720 MHz maser
and present the results of calculations that show the effects of gas
temperature, molecular density and dust radiation on the maser intensity. In
section III we analyze the constraints the pumping imposes on the type of shock
and present a shock model that is in good agreement with the 1720 MHz maser
observations. Section IV contains our conclusions.

\section{ Collisional Pumping of the 1720 MHz Maser }
At the low temperatures in the 1720 maser region, most of the OH molecules
populate the four hyperfine levels of the ground state \ulp($J=3/2$) (see
figure 1). Excitations to the higher rotational levels followed by radiative
cascades back to the ground reshuffle the molecules among the ground-state
hyperfine levels, leading to inversion of one of the satellite lines when the
final cascade is optically thick. When that final decay is from the first
excited state, \ulp($J=5/2$), the 1720 MHz transition is inverted. However,
when the final decay is from the next excited state, \urp($J=1/2$), the 1612
Mhz line is inverted instead. As a result, the 1720 MHz inversion occurs only
if the final cascade back to the ground is dominated by optically thick decays
from the first excited state. This condition is met only when excitations are
by collisions at $T$ \la\ 200 K, so that \ulp($J=5/2$) is preferentially
populated, and \NOH\ \la\ \E{17} \cs, so that decays from \urp($J=1/2$) are
optically thin; when \NOH\ is increased sufficiently, the inversion switches to
the 1612 Mhz line. These basic principles have been established long ago (E76)
and verified in numerous observations, most notably by van Langevelde et al.
(1995).

The mechanism responsible for the 1720 MHz inversion arises from the basic
energy level structure of OH. In particular, it involves simple level counting
and is largely independent of the specific form of collision rates; inversion
should occur as long as these rates do not exhibit peculiar selection rules.
Due to the lack of calculated collision rates, the E76 study employed hard
sphere rates in which all downward collisions have equal strength. The
calculation of actual OH collision cross sections is difficult, and the initial
attempts were plagued by serious errors. These mistakes, reviewed by Flower
(1990), sometimes led to completely opposite conclusions concerning the pumping
of the various OH maser lines. However, the latest collision rates calculated
by Offer et al.\ (1994) for the 24 hyperfine levels shown in figure 1 are more
accurate than earlier ones and are also in reasonable agreement with
experiment.

The present work utilizes the Offer et al.\ (1994) collision rates to update
the original E76 calculations. Similar to that study, the level populations are
calculated in the escape probability formalism and here we employ the
expression developed by Capriotti (1965) for a quiescent slab. We have also
investigated the effect of dust continuum radiation on the maser strength. The
dust intensity is computed from the approximation
\begin{equation}
                  I_\nu = [1 - e^{-\tau_{\nu,d}}] B_\nu(T_d)
\end{equation}
where $B_\nu(T_d)$ is the Planck function of the dust temperature $T_d$ and
$\tau_{\nu,d}$ is the dust optical depth with a spectral shape following the
standard ISM dust cross section (Draine \& Lee 1984). The calculation of the OH
level populations is complicated by the phenomenon of line overlap, which
occurs when the frequencies of two or more spectral lines are sufficiently
close that photons emitted by one can be absorbed by another. This affects the
escape of the photons and the level populations. We have incorporated the
effects of overlap due to thermal motions with the aid of the method of Lockett
\& Elitzur (1989).

\subsection{Results}
We compute the populations of the 24 levels shown in figure 1 for a variety of
physical parameters, assuming thermal linewidths. From the level populations we
find the maser optical depth $\tau_m$ in the direction of shock propagation.
The largest maser amplification will occur in the direction of largest velocity
coherence and will tend to be perpendicular to the shock motion, similar to the
case of water masers in star-forming regions (Elitzur, Hollenbach \& McKee
1989). Our results are presented in figures (2)--(4), which display the effects
of temperature, density and external radiation field on the maser. The figures
show how each property affects the variation of maser optical depth $\tau_m$
with OH column density \NOH. Figure (2) shows the effect of gas temperature.
Strong maser emission constrains the temperature to be between about 50 and 125
K, with maximum inversion occurring for $T$ \about\ 75 K. It is interesting to
note that the upper limit on the temperature is affected by line overlap.
Without overlap, significant inversion would persist beyond 125 K, declining
only when $T$ exceeds 150 K.

Figure (3) shows the effect of molecular density. The maser cannot operate at
high densities since quenching by collisions occurs at densities above about
3\x\E5 \cc. Although the inversion persists as the density decreases, very low
densities are not compatible with realistic maser gains, $\tau_m$ \ga\ 2, which
require \NOH\ \ga\ 4\x\E{15} \cs. Since the OH abundance relative to H$_2$ is
probably no more than \about\ \E{-5}, the corresponding overall column density
is \ga\ 4\x\E{20} \cs. Both observations and model calculations discussed below
limit the OH shell width to less than \about\ 2\x\E{16} cm, thus the overall
density in the region must exceed \about\ \E4 \cc. Our calculations therefore
suggest that the range of molecular densities for observable maser emission
most likely lie between $10^{4}$ and 5\x\E5 \cc.

Figure (4) presents the effect of the dust continuum radiation on the maser
strength, showing that dust at a temperature $T_d$ = 50 K has little impact on
the maser effect. However, at $T_d$ = 100 K the dust radiation diminishes
significantly the maser inversion even with a minimal amount of dust. SNR's
often show strong IR continuum emission which would tend to reduce the 1720 MHz
maser inversion. However, observations indicate relatively low dust emission
and temperatures in the maser sources IC443 and 3C391. Burton et al. (1990)
find that the IR emission from IC443 is dominated by line emission rather than
by dust continuum radiation. When line emission is taken into account, a dust
temperature of \about\ 45 K is found (van Dishoeck et al.\ 1993). Reach \& Rho
(1996) have recently measured the IR continuum emission from 3C391 using ISO
and find a dust temperature of 50 K. Figure 4 shows that in these sources the
dust radiation will not seriously reduce the maser inversion. Furthermore, the
galactic center masers observed by Yusef-Zadeh et al. (1996) are located
outside the region of intense IR emission (Mezger, Duschl, \& Zylka 1996).

The results presented here are similar to those derived in E76 with hard sphere
collisions, corroborating the conclusion of that study that the 1720 MHz
inversion reflects the basic OH energy level structure and is largely
independent of the exact form of collision rates. Because of the severe
limitation on OH column density, the maser optical depth is never large.
Indeed, even under optimal pumping conditions the 1720 MHz maser barely reaches
saturation for radiation propagating along the shock direction; when $\tau_m$
is increased to the point that amplified maser intensity is at saturation
level, \NOH\ is already so large that the inversion switches to the 1612 MHz
maser. However, significant brightness temperatures can be produced for
radiation propagating parallel to the shock front with amplification optical
depth $a\tau_m$, where $a$ is an aspect ratio.  For \NOH\ \about\ \E{16} \cs,
modest aspect ratios of only 3--4 suffice to produce brightness temperatures in
the range \E{8}--\E{10} K.

Detailed OH pumping calculations based on the Offer et al. (1994) rates were
also performed by Pavlakis \& Kylafis (1996; PK96 hereafter), and our results
are in general agreement with theirs. The Offer et al.\ calculations were the
first to include collisions with both ortho- and para-H$_2$, and PK96 found
that collisions with para-H$_2$ can eliminate the 1720 MHz inversion. They
suggested, therefore, that the 1720 MHz maser might be used as a tool to
determine the ortho-to-para ratio of molecular hydrogen. Our calculations
support the PK96 finding that para-H$_2$ collisions do not lead to 1720 MHz
inversion. However, we conclude that this inversion may not provide a useful
indicator of the H$_2$ ortho-to-para ratio. First, an examination of the cross
sections shows that the dramatic difference in maser efficiency between the
ortho- and para-H$_{2}$ collisions reflects a minor peculiarity of the cross
sections. Para collisions deplete the maser upper level more than the lower
level while ortho collisions deplete both at roughly the same rate. This
disparity arises mostly because of a single collision rate: the rate from
$\ulp(J = 5/2, F = 3^+)$ to the maser's lower level is more than 100 times
smaller than that to the upper level for the para collisions. Merely setting
these two rates equal to each other, as they are for the ortho collisions,
produces a 1720 MHz inversion also for para-H$_{2}$. Thus the lack of a 1720
MHz maser depends sensitively on just a couple of cross sections. Although the
latest rates are the most exact and complete that have yet been calculated, the
history of OH collision rate calculations suggests that caution should be
exercised in reaching conclusions based on selective collision rates.

A second deficiency of the 1720 MHz maser as a diagnostic of the ortho to para
ratio is that it takes just a small amount of ortho to produce the maser since
the relevant ortho collision rates are much larger than the para rates. The
equilibrium ratio of ground state ortho to para H$_2$ is 9\,exp($-170/T$).
However, the ratio is predicted to be higher than the equilibrium value at low
temperatures (Le Bourlot 1991), indeed, observations of protostellar outflows
yield a ratio of $3\pm 0.4 $ (Smith, Davis \& Lioure 1997). It is believed that
H$_2$ forms on grains at an ortho to para ratio of 3:1, and that radiative
decays preserve this ratio as the gas cools down. The results presented in
figures (2)--(4) used an ortho to para ratio of 3:1.  Strong 1720 MHz maser
emission would persist as long as this ratio is larger than one. Although this
ratio will probably influence the strength of the 1720 MHz maser, it would be
difficult to estimate its value from the maser intensity alone.

\section{The Maser Region --- C-Shock Diagnostic}

Our pumping calculations predict that the 1720 MHz maser should arise under the
following conditions: $T$ \about\ 50--125 K, $n$ \about\ \E5 \cc, \NOH\ \about\
\E{16}--\E{17} \cs. These parameters are in excellent agreement with the
observations (\S1), which clearly establish that the 1720 MHz masers are
produced at the shock interface of a SNR with a molecular cloud. It is also
significant that these conditions do not result in production of the other
ground state masers. As discussed earlier (\S2), the 1612 MHz maser requires a
larger OH column for its creation while the main line masers require high dust
temperatures (Elitzur 1992), which would destroy the 1720 MHz
maser\footnote{This is perhaps the reason why the 1720 MHz maser is the only
one of the ground state OH masers not observed in the envelopes of late-type
stars.}. It is also important that a recent search did not detect the 22 GHz water maser
in association with the 1720 MHz maser (Claussen et al. 1998).
This non-detection is significant, since observable 22 GHz water maser emission
is not expected for the densities and temperatures present in these regions.
We now model the environment that can generate these conditions, and
show that it requires a C-type shock. Thus the production of the 1720 MHz maser
can be used as a diagnostic of the shock structure.

Our conclusion that the shock must be C-type is based solely on the physical
conditions necessary for 1720 MHz maser action. Because of the rapid cooling of
the shocked material, the time spent in the necessary temperature range is so
short that the accumulated column density is much too small to produce
observable maser emission. We will show that ambipolar diffusion heating in
C-shocks can overcome this rapid cooling, but that no equivalent heat source
exists for J-shocks. Consider the material leaving a shock front, where it was
heated to a high temperature. As this hot material streams away, it both cools
and slows down so that its temperature $T$, velocity $v$, and density $n$ are
unique functions of distance $l$ from the front. Assuming an elevated OH
abundance of \E{-5} gives a lower limit of
\about\ \E{21} \cs\ for the H$_2$ column density of the maser, and the shock
must be able to produce such a column for the shocked material as it cools down
from $T_h$ \about\ 125 K to $T_l$ \about\ 50 K. From mass conservation, the
column contained between these two temperatures is (Elitzur 1980)
\begin{equation}
        N_{\rm H_2}=\int n dl = \int n v dt
                = n_0v_s\int_{T_l}^{T_h} t_{\rm cool}{dT \over T}
\end{equation}
where $n_0$ is the pre-shock density, $v_s$ is the shock velocity and $t_{\rm
cool} = |d\ln T/dt|^{-1}$ is the cooling timescale (= $\case{5}{2}kT/\Lambda$,
where $\Lambda$ is the net energy loss rate per particle). This relation is a
direct consequence of mass conservation and applies to all shocks. Its
evaluation requires an estimate of $t_{\rm cool}$, i.e., $\Lambda$. Assume
first that there is no heating in the shocked region, as is the case in
standard J-shocks, then $\Lambda$ is determined by the cooling processes only.
Detailed calculations show that the dominant coolant in the relevant
temperature region is O{\small I}, through its $^3P$ fine structure transitions
(e.g.\ Hollenbach \& McKee 1989), and simple estimates show that $\Lambda
\simeq$ 6\x\E{-25}\,$T$ erg\,s$^{-1}$ per H$_2$ in this case. This gives a
temperature-independent cooling timescale \about\ 6\x\E8 s, an upper limit on
$t_{\rm cool}$ since other coolants may also contribute.  Thus the previous
equation becomes
\begin{equation} \label{column}
        N_{\rm H_2} = n_0v_s t_{\rm cool}\ln{T_h\over T_l}\
                    \la\ 6\x\E{18}\, n_{0,4}v_6\ \cs,
\end{equation}
where $n_{0,4} = n_0/(\E4\,\cc)$ and $v_6 = v_s/(\E6\, \rm cm\,s^{-1})$. To
support the 1720 MHz maser this column must be \about\ \E{21} \cs, therefore
the ambient density and shock velocity are constrained by
\begin{equation}\label{constraint1}
                        n_{0,4}v_6\ \ga\ 100.
\end{equation}
Now, in a J-shock, to a good degree of approximation the ram pressure ($\propto
n_0v_s^2$) is equal to the thermal pressure of the shocked material (Elitzur
1980). Since the pumping requirements place upper limits on both the density
($n$ \la\ 5\x\E5 \cc) and temperature ($T$ \la\ 125 K) of the maser region, it
follows that its thermal pressure is bounded by $nkT$ \la\ 6.25\x\E7k erg~\cc.
Therefore, pressure balance across a J-shock yields
\begin{equation}
                        n_{0,4}v_6^2\ \la\ 0.3,
\end{equation}
another constraint on $n_0$ and $v_s$. The last two constraints can be obeyed
simultaneously only if $v_s$ \la\ 3\x\E3 cm\,s$^{-1}$ and $n_0$ \ga\ 3\x\E8
\cc, and these two bounds are both unphysical. The upper bound on the shock
velocity is two orders of magnitude less than the speed of sound, and the {\em
lower} limit on the {\em pre-shock} density exceeds the {\em upper} limit on
the {\em post-shock} maser density by more than three orders of magnitude.
Theses difficulties persist for J-shocks even when the post-shock pressure is
dominated not by the thermal motions but rather by a frozen-in magnetic field.
Assuming flux freezing and an ambient field $B = \E{-6}(n_0/1~\cc)^{1/2}$\,G,
pressure-balance across the shock yields $n_{0,4}v_6 = 0.1 n_4$, where $n =
\E4n_4$ \cc\ is the post-shock density (e.g.\ Hollenbach \& McKee 1989).
Together with equation (\ref{constraint1}) this implies $n$ \ga\ \E7 \cc, a
lower limit exceeding the upper bound on maser density by more than an order of
magnitude.

J-shocks cannot produce the relatively high column densities necessary for
observable maser emission at the relatively low thermal pressures dictated by
the pumping requirements of the 1720 MHz maser. The failure can be traced
directly to the low column densities that these shocks produce in the relevant
temperature region. From equation \ref{column}, typical ambient densities and
shock velocities produce column densities that are more than two orders of
magnitude smaller than required. The problem is that $t_{\rm cool}$ is so
short, since the cooling is so fast, and can be circumvented only if some
heating existed to counter the cooling and maintain the temperature in the
50--125 K range for a longer period, extending the column. One possible heat
source in J-shocks is the heat of H$_2$ formation on grains, which can be
tapped if the shock is sufficiently fast to first dissociate all molecules
(Hollenbach, McKee, \& Chernoff 1987).  Indeed, this is the heating mechanism
invoked to create the right conditions for water masers in dissociative
J-shocks with pre-shock densities \ga\ 3\x\E6 \cc\ (Elitzur et al.\ 1989).
However, since this mechanism involves collisional de-excitation of H$_2$
vibrational states, its efficiency deteriorates so rapidly at lower densities
and temperatures that it becomes inoperable at the conditions required here.
Another possibility is radiative heating.  One common method is to first
radiatively heat the dust, transferring the heat to the gas by collisions. This
mechanism is not feasible here because the radiation from the heated dust would
destroy the maser inversion. Another radiative heat source could be ionization
by the SNR x-rays, since a significant fraction of the ionization energy goes
into heating. Maloney, Hollenbach \& Tielens (1996) estimate that 30--40\% of
the ionization energy goes into heating the gas if it is primarily molecular.
From the expressions they provide, the x-ray heating rate for the conditions
considered here is only $\sim \E{-27}$ erg\,s$^{-1}$ per H$_2$, about 4 orders
of magnitude less than the cooling rate and thus a negligible heating source.

Heating of the shocked material can be provided by ambipolar diffusion. This
mechanism occurs when flux-freezing breaks down and the shock then switches to
C-type. In the parameter regime relevant here, especially the magnetic fields
inferred from the polarization measurements, this is the expected shock type.
Indeed, in their analysis of line emission from IC443, Burton et al.\ (1990)
concluded that C-shocks are present in that source. Furthermore, Reach and Rho
(1998a) have recently found that the brightness of the CO lines they observed
in the 3C391 maser region is consistent with C-type molecular shocks with $10^4
< n_0 < 10^5$ cm$^{-3}$ and $10 < v_s < 50$ \kms.

Detailed C-shock calculations were performed by Draine, Roberge and Dalgarno
(1983, DRD hereafter).  They present results for a number of models that
bracket the general requirements for 1720 MHz maser operation. In particular,
their intermediate density model has $n_0$ = \E4 \cc\ and $v_s$ = 25 \kms. This
shock develops a region with the appropriate temperature (50--125 K) and
density (\about\ \E5 \cc) that extends over a distance of \about\ \E{16} cm,
i.e., a column density of \about\ \E{21} \cs. These are just the parameters
required for the 1720 MHz maser. Unfortunately, this potential maser region has
virtually no OH; as is well known, molecular shocks channel into H$_2$O, rather
than OH, essentially all the oxygen not in CO, and this is the case also with
the DRD model. However, subsequent photodissociation, which was not included in
the DRD calculations, can convert the water into OH. Such dissociation could be
produced by the external UV radiation field as in the case of shocks around
HII/OH regions (Elitzur and de Jong 1978). The relevant photodissociation rates
were listed most recently by Roberge et al. (1991)
\begin{equation}
 R({\rm H_2O}) = 3.2\x10^{-10}\chi\exp(-2.5\tau_V)\ {\rm sec^{-1}}, \qquad
 R({\rm OH})   = 1.9\x10^{-10}\chi\exp(-2.5\tau_V)\ {\rm sec^{-1}}.
\end{equation}
Here $\chi$ is the local enhancement of the standard interstellar UV field
(Draine 1978) and $\tau_V$ is the source's optical depth at visual wavelengths.
With these rates we can solve analytically for the time dependent OH and H$_2$O
abundances of any parcel of gas after its exposure to UV radiation. Note that
time $t$ enters into the equations only in the combination
$t\chi\exp(-2.5\tau_V)$. The solution for a mixture that is initially purely
H$_2$O is displayed in figure (5), which shows the OH abundance as a fraction
of the gaseous oxygen not in CO.

Another dissociation mechanism, recently advocated by Wardle et al.\ (1998),
involves the secondary UV photons produced after X-ray absorption. With
expressions from Maloney et al.\ (1996), this water dissociation rate is
\begin{equation}
 R({\rm H_2O}) \simeq 5\x\E{-12}\,{L_{36}\over R_{pc}^2 N_{22}}\ {\rm sec^{-1}}
\end{equation}
where $\E{36}L_{36}$ erg s$^{-1}$ is the SNR X-ray luminosity, $R_{pc}$~pc is
the SNR radius and $\E{22}N_{22}$ \cs\ is the hydrogen column attenuating the
X-ray flux. The three SNRs IC443, W44 and 3C391 have $L_{36}$ \about\ .1--2,
$R_{pc}$ \about\ 10 and $N_{22}$ \about\ 0.1--1 (Charles \& Seward 1995; Rho et
al.\ 1994; Rho \& Petre 1996). These parameters yield a water dissociation rate
of \about\ \E{-13} sec$^{-1}$, about 1000 times smaller than that due to the
standard ISM field. Therefore this mechanism can be neglected, unless the x-ray
flux has been underestimated by more than a factor of 50.

The width of the maser region in the DRD model is \about\ \E{16} cm.  With a
shock speed of 25 \kms, the corresponding crossing time is \about\ 125 years,
same as the time of the peak evident in figure 5.  Thus this figure suggests
that maximum OH abundance would occur for $\chi\exp(-2.5\tau_V) \sim 1$, the
strength of the unattenuated standard interstellar field.  However, radiation
this intense would significantly increase the electron abundance and could
drastically affect the shock structure to the point that it is no longer a
C-shock. DRD note that a modest change in the pre-shock electron density will
not significantly affect the structure of their \E4 \cc\ C-shock because most
of the momentum transfer is via charged dust grains. However, the standard,
unshielded interstellar field would result in an increase of the electron
abundance from \E{-7} to about \E{-5}, and such a large increase would have a
significant effect on the shock structure. Thus there are two competing
constraints on the strength of the UV field: it must be strong enough to create
sufficient OH, yet not so strong as to increase the electron abundance too
much. A field 100 times weaker than the standard interstellar field appears to
be a reasonable choice. It will increase the electron abundance by no more than
an order of magnitude while producing the required amount of OH.  With
$\chi\exp(-2.5\tau_V) \sim \E{-2}$, the time axis in figure 5 is stretched so
that roughly 10\% of the water is converted to OH in the maser region and the
OH abundance is about 2\x\E{-5}. These results are in good agreement with the
recent observations of Reach and Rho (1998b) who detected both OH and H$_2$O at
an abundance ratio of $\sim$ 1:15 in a shocked region in 3C391. Although more
accurate estimates of the OH abundance will require shock calculations that
directly include photodissociation effects, this analysis shows that water
photodissociation can lead to the necessary OH column density without serious
effects on the shock structure.

Finally, the ambipolar diffusion process is driven by magnetic field gradients.
In a C-shock these gradients are generated by the shock compression.
Additionally, gradients can be produced by field curvature, an attractive
possibility given the apparent maser location at the edge of an expanding SNR.
A gradient characterized by length scale \about\ \E{17} cm would produce
ambipolar diffusion whose heating rate is similar to that of the DRD C-shock
model. Such a gradient, which could be reasonably expected at the interface of
clumps overtaken by the SNR shell, would augment the C-shock heating effect
even if the ionization were higher than what DRD assumed in their calculation.

\section{Discussion}

For many years the 1720 MHz transition was the least observed of the OH ground
state maser lines. Our work shows that the inherent weakness and rarity of this
maser reflect a combination of independent factors. First, because of the
severe limitation on OH column density, the 1720 MMz maser optical depth is
never large. Even when produced under optimal conditions, appreciable maser
amplification would require radiation propagation parallel to the shock front
and sizeable aspect ratios. Second, the production of this maser places tight
constraints on the physical environment. Furthermore, the particular
combination of physical parameters necessary for this maser is quite uncommon
and only C-shocks seem capable of producing it. Third, even C-shocks are
incapable of generating the optimal conditions for this maser. The OH columns
produced by these shocks correspond to $\tau_m$ smaller than its potential peak
value by more than a factor of 2, a significant factor in light of the steep
dependence of maser emission on optical depth. Thus our work provides a natural
explanation for the low detection rate of \about\ 10\% found by Green et al.
(1997).

While the tight constraints on the physical conditions needed for its
production make for a weak 1720 MHz maser, they also make it a powerful probe
of the region where it is formed. Our most significant result is that the
detection of the 1720 MHz maser rules out the possibility of a J-shock; only
C-shocks can produce the conditions required for this maser. This conclusion,
too, is supported by the detection of high maser polarization and the magnetic
fields it implies. Because of the wide margin by which J-shocks fail to produce
it, the 1720 MHz maser could be one of the best discriminators of the shock
type, a reliable indicator of the presence of a C-shock.

\acknowledgments We thank E. van Dishoeck for sending us the tabulations of the
Offer et al.\ collision rates, and D. Frail, D. Hollenbach and W. Reach for
most useful comments on the original manuscript. The partial support of NASA
grant NAG5-7031 is gratefully acknowledged. After completion of this work we
have learned of a study by Wardle, Yusef-Zadeh \& Geballe (1998) that reached
similar conclusions regarding the role of C-shocks for these masers.

\appendix
\section*{Appendix: Scaling}

In the absence of external radiation, the level populations are determined by
the following parameters: overall density $n$, $\NOH/\Delta v$ and the
temperature. When the pumping is collisional and all the relevant lines are
optically thick, the first two parameters enter only in the combination
\begin{equation}
                 \xi = {n\NOH\over\Delta v},
\end{equation}
a scaling parameter that proved useful in the analysis of water maser pumping
(Elitzur et al.\ 1989). Since the 1720 MHz maser, too, is collisionally pumped,
we investigated the scaling properties of its inversion efficiency $\eta =
(n_{u}-n_{l})/(n_{u}+n_{l})$. We find that to a good degree of approximation,
$\eta$ is indeed a function only of $T$ and $\xi$, i.e., $n$-independent, when
$\xi$ \ga\ \E{21} cm$^{-5}$/\kms\ and $n \le$ 3\x\E5 \cc. At smaller values of
$\xi$ we find that $\eta$ remains independent of $n$ but $\xi$ is replaced by
$\NOH/\Delta v$ as the relevant scaling variable. Both scaling properties can
be verified analytically with the aid of a simple 8-level model of the first
two rotation states, employing hard-sphere collision rates. In this model the
inversion efficiency obeys
\begin{equation}
        \eta \propto \frac{7R_{8,4}}{R_{8,4}+4C} +
                     \frac{5R_{7,4}}{R_{7,3}+R_{7,4}+4C} -
                     \frac{5R_{5,1}}{R_{5,1}+R_{5,2}+4C}
\end{equation}
where level numbering is in order of increasing energy (see figure 1), $R_{ij}
= A_{ij}\beta_{ij}$ is the radiative decay rate and $C$ is the (uniform) downward
collision rate. Note that $C$ depends only on $T$ and $n$ while $R_{ij}$
depends only on $\NOH/\Delta v$. At low densities and column densities $C <
R_{ij}$, therefore $\eta$ depends only on $R_{ij}$, i.e., $\NOH/\Delta v$. The
dependence on $\NOH/\Delta v$ persists as long as some of the radiative
transitions remain optically thin. As the column density increases, the
radiative decay rates become smaller, due to photon trapping, and downward
collisions begin to dominate. Eventually all the relevant optical depths become
large and the maser efficiency becomes a function of $R_{ij}/C$, i.e., it
scales with $\xi$.

Thanks to scaling, the numerical results presented here can be extended to
other densities. However, because of its small phase space, the usefulness of
scaling for the 1720 MHz maser is more limited than for water masers.

\bigskip

\begin{figure}[h]          
 \centering \leavevmode \epsfxsize = 0.7\hsize \epsfclipon \epsfbox{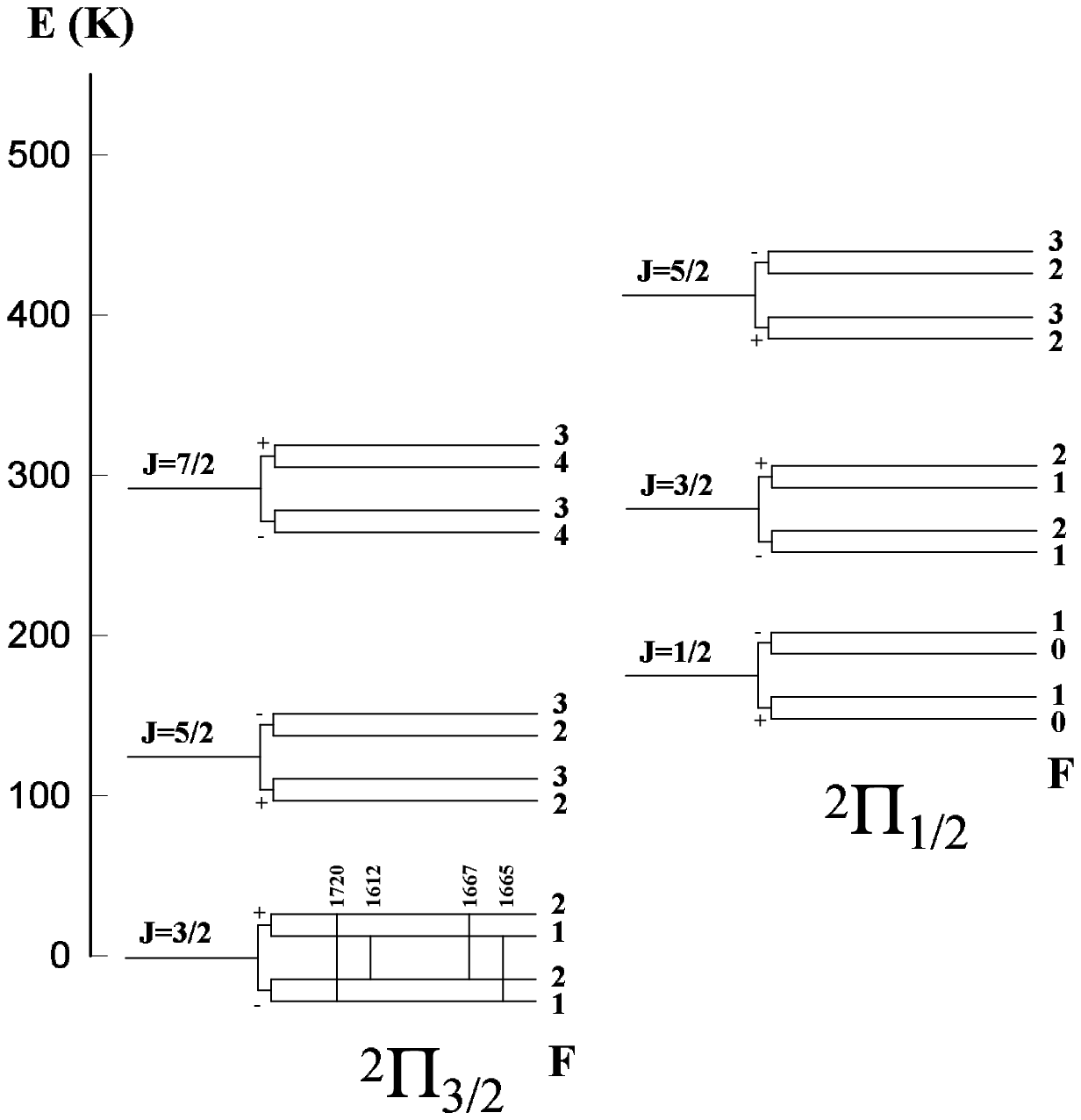}
\figcaption[fig1.ps]{Hyperfine-rotation levels of OH used in our calculations.}
\end{figure}
\begin{figure}          
 \centering \leavevmode \epsfxsize = 0.7\hsize \epsfclipon \epsfbox{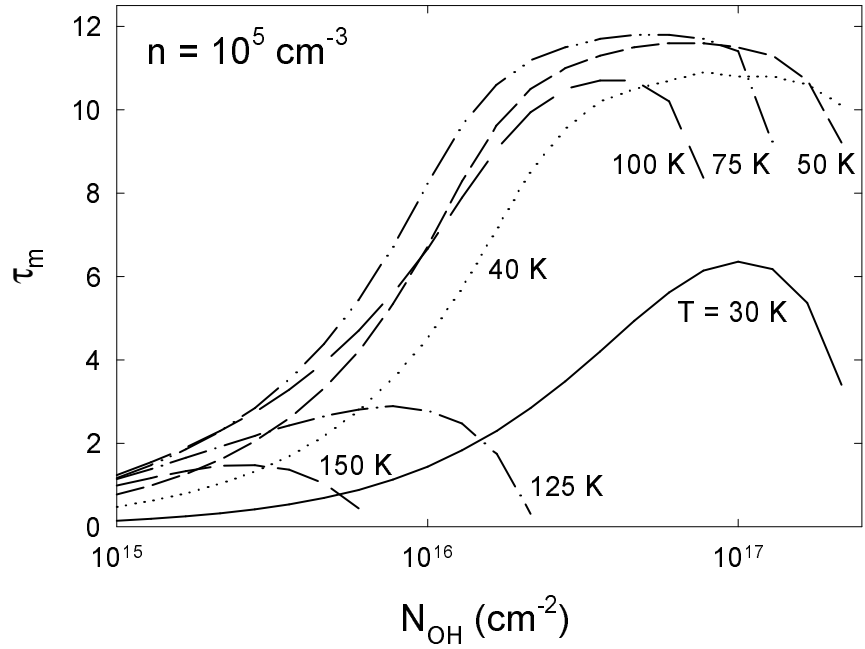}
\figcaption[fig2.ps]{Maser optical depth as a function of OH column density for
various temperatures, as marked, at density $n$ = \E5 \cc.}
\end{figure}
\begin{figure}          
 \centering \leavevmode \epsfxsize = 0.7\hsize \epsfclipon \epsfbox{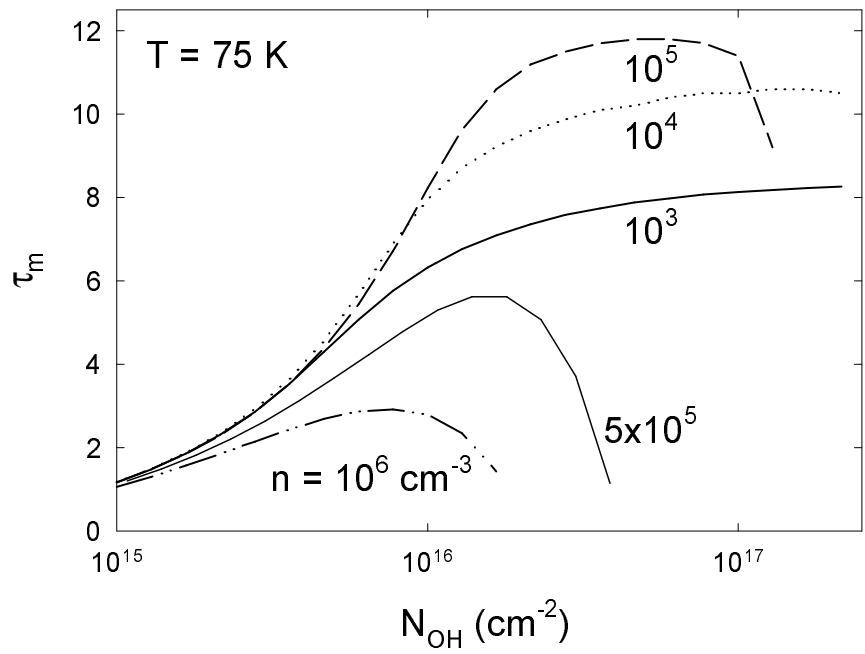}
\figcaption[fig3.ps]{Maser optical depth as a function of OH column density for
various densities, as marked, at temperature $T$ = 75 K.}
\end{figure}
\begin{figure}          
 \centering \leavevmode \epsfxsize = 0.7\hsize \epsfclipon \epsfbox{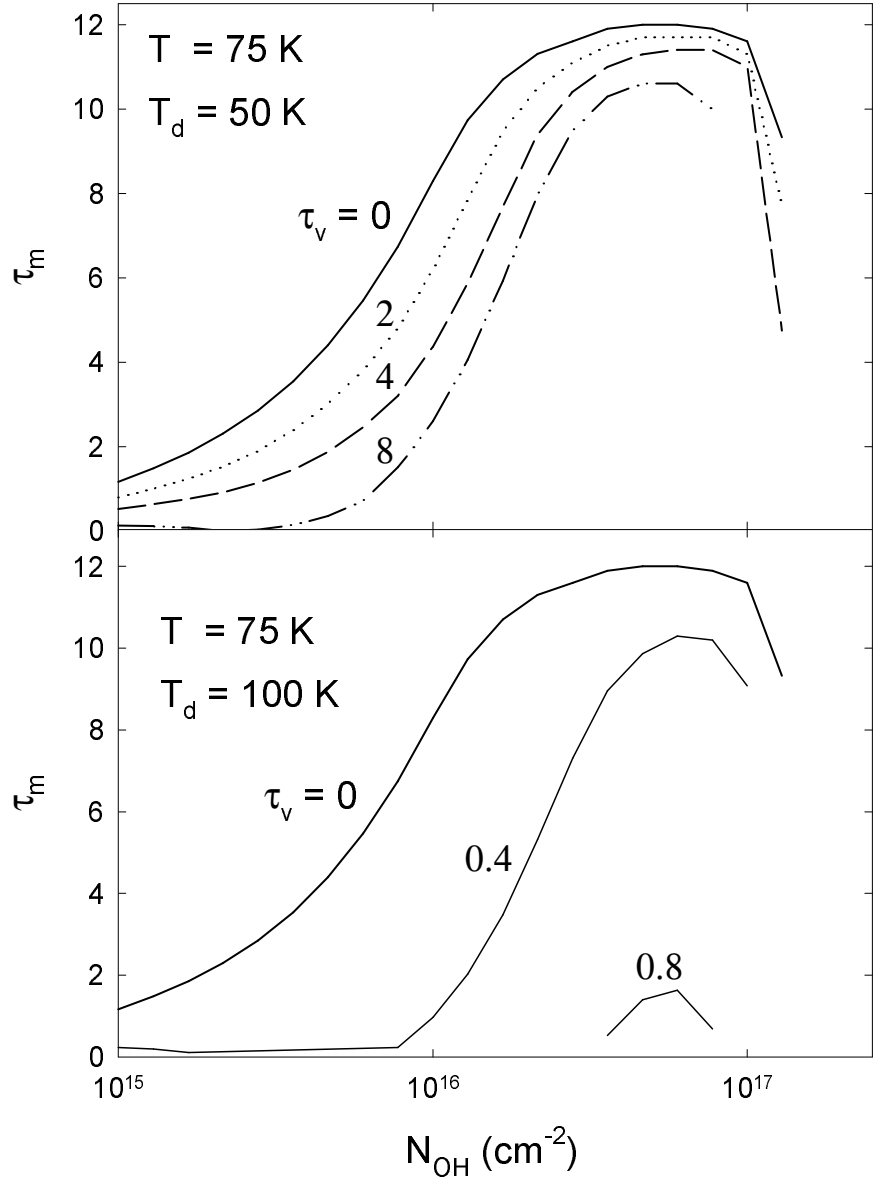}
\figcaption[fig4.ps]{Effect of dust radiation on maser inversion: maser optical
depth as a function of OH column density for various values of $\tau_V$, the
dust optical depth at visual, as marked. The dust temperature is 50 K in the
top panel, 100 K in the bottom panel. In both panels, the gas has a density of
\E5 \cc\ and temperature 75 K. The plots for $\tau_V$ = 0 are the same as the
corresponding ones in figures (2) and (3) and are included for reference, to
show the inversion in the absence of dust.}
\end{figure}
\begin{figure}          
 \centering \leavevmode \epsfxsize = 0.7\hsize \epsfclipon \epsfbox{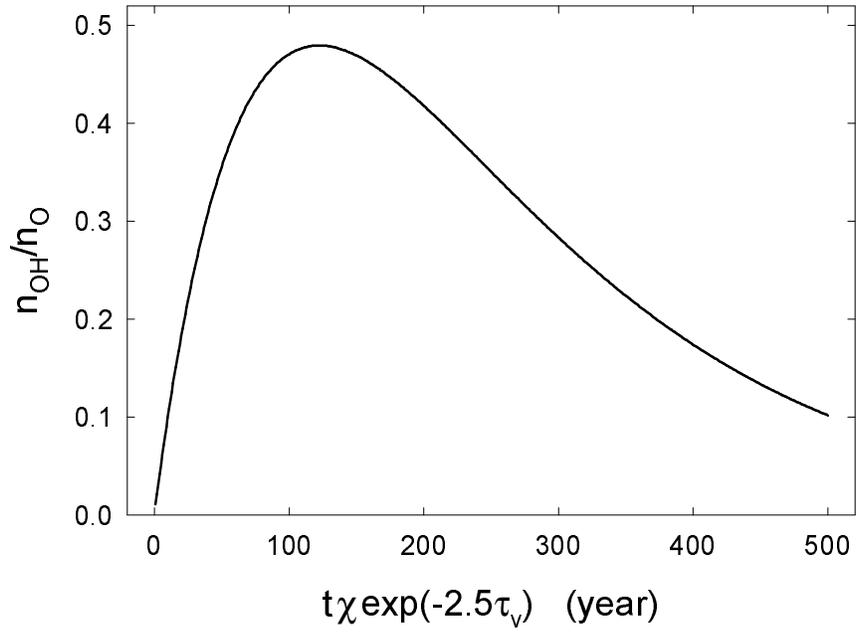}
\figcaption[fig5.ps]{Fraction of oxygen in OH as a function of time for gas
exposed to UV radiation. Initially, all oxygen is in H$_2$O. $\chi$ is the
enhancement of the local radiation field over its standard ISM value and
$\tau_v$ is the cloud's visual extinction. The scaling properties of the time
axis directly reflect the form of the photodissociation rates (eq.\ 6).}
\end{figure}

\end{document}